\documentclass[a4paper,11pt]{article}

\usepackage{pos}

\usepackage{lipsum} 

\title{DISCO: An optical instrument to calibrate neutrino detection in complex media}

\ShortTitle{DISCO}

\author{Carsten Rott$^{1,2,*}$}
\author{Segev BenZvi$^{3}$}
\author{Mike DuVernois$^{4}$}
\author{Kenneth Golden$^{5}$}
\author{Benjamin Jones$^{6}$}
\author{Christoph Toennis$^{2}$}

\emailAdd{rott@physics.utah.edu}

\abstract{

We present a conceptual design of a high-performance camera system with applications to neutrino detectors, deep sea exploration, and glaciology. The design combines ultra-sensitive cameras with a number of well-calibrated light sources enclosed in a pressure vessel. The instrument will be capable of withstanding extreme environments such as those encountered in Antarctica or the deep ocean, and be deployable as a standalone system that can be retrieved for deep-sea exploration or glaciology. The camera system is designed to be replicated and deployed in multiple detectors, requiring only modest modifications from one detector to another. The instrument combines a number of capabilities essential for neutrino detector calibrations, including characterization of the scattering and absorption properties of the optical medium, measurement of geometries via photogrammetry, and detector surveillance. The ability to deploy the instrument at different detector sites also offers opportunities for cross-calibration efforts. We present the conceptual design of the instrument and describe plans to produce a prototype.

\vspace{4mm}
{\bfseries Corresponding authors:}
Carsten Rott$^{1,2,*}$\\
{$^{1}$ \itshape Department of Physics and Astronomy, University of Utah, Salt Lake City, UT 84112, USA}\\
{$^{2}$ \itshape Department of Physics, Sungkyunkwan University, Suwon, 16419, South Korea}\\
{$^{3}$ \itshape Department of Physics and Astronomy, University of Rochester, Rochester, NY 14627, USA}\\
{$^{4}$ \itshape Department of Physics and Wisconsin IceCube Particle Astrophysics Center, University of Wisconsin–Madison, Madison, WI 53706, USA} \\
{$^{5}$ \itshape Department of Mathematics, University of Utah, Salt Lake City, UT 84112, USA} \\
{$^{6}$ \itshape Department of Physics, University of Texas at Arlington, 502 Yates St., Science Hall Rm 108, Box 19059, Arlington, TX 76019, USA}
\\[4mm]
$^*$ Presenter

\ConferenceLogo{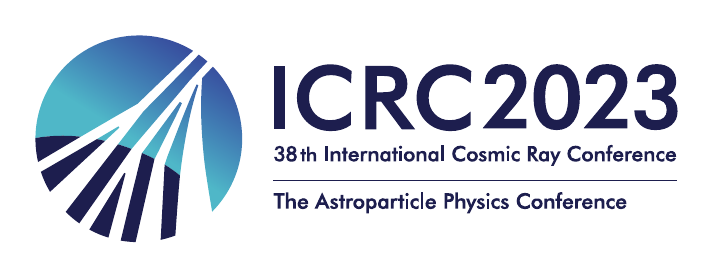}

\FullConference{The 38th International Cosmic Ray Conference (ICRC2023)\\ 26 July -- 3 August, 2023\\ Nagoya, Japan}
}

\begin{document}

\maketitle

\section{Introduction}\label{sec1}
\vspace{-.2cm}

A new era in 
astroparticle physics has begun with the discovery of high-energy astrophysical neutrinos, a coincidence observation of energetic neutrinos and $\gamma$-rays from a blazer~\cite{IC:TXSblazar}, 
evidence for neutrino emission from NGC1068~\cite{IceCube:2022der}
and from the Milky Way by the IceCube Observatory. 
Further progress in this emerging field is fueled by multiple large volume neutrino detectors now operating, all with funded efforts for expansions:
IceCube~\cite{IceCube:PeV} with its Upgrade~\cite{Ishihara:2019uL} at the geographic South Pole, the Mediterranean ANTARES 
with KM3NeT~\cite{Adri_n_Mart_nez_2016},  
and Lake BAIKAL~\cite{BAIKAL} with the GVD extension. Proposed large detectors include P-ONE~\cite{Rea:2021o3} in the Pacific, IceCube-Gen2~\cite{IceCubeGen2}, and TRIDENT.
 
Neutrino telescopes utilize naturally occurring media with good optical properties. Gigaton scale volumes are instrumented with thousands of optical sensor modules to create a grid for neutrino detection via the Cherenkov light associated with high-energy neutrino interactions. 
Exploiting the physics capabilities of these detectors requires precise modelling of optical properties of the detection medium. The medium is naturally occurring, so its properties are as-found and may be non-uniform (in the case of ice) or time-dependent (in the case of water). Astronomical observations  
rely on precise characterization of the in-medium properties; conversely, imperfectly characterized ice and water properties remain a major component of their systematic uncertainty budgets. 

Neutrino telescopes have used pulsed flasher LEDs for calibration in the past, inferring optical properties of the medium from detected pulse shapes~\cite{ANTARES:2018tce,ICdetector}. 

In these systems, emitter and receiver are fixed at particular points in the array with light traveling large distances through varying optical conditions, inhibiting the extraction of truly local properties. Logging devices with deployment plans similar to ours have also been used, but with optical systems and calibration insufficient to extract absolute properties or to disentangle effects of absorption and scattering. 
The use of optical images to understand complex media at neutrino telescopes was demonstrated with the deployment of a camera system to monitor the freeze-in process in the IceCube drillholes~\cite{ICdetector,refId0}. Two cameras, accompanied by laser diodes, observed several unexpected features, including the formation of bubbly ice around detector modules. Cameras are an integral part of the IceCube Upgrade calibration plan. Prototypes of the system were successfully deployed at P-ONE, and an ice core drill hole near the geographic South Pole~\cite{Toennis:SPICECAM2021,Pollmann:Lumi2021}. 

Polar sea ice forms one of the key components of Earth's climate system that is most impacted by planetary warming. As a material, it is a multiscale, polycrystalline
composite of pure ice with brine, air, and particulate inclusions. 
Moreover, the sea ice interior hosts extensive communities of 
microbial extremophiles like algae, that serve as a 
critical lynch-pin in the marine food web. 
Improving projections of the health of  
polar ecosystems in the 
face of precipitous sea ice losses, requires advances in 
understanding how algal biomass distributions are correlated with 
local ice conditions.
Wide-ranging surveys of these internal properties over large expanses of sea ice with portable instruments are not presently practical, with much present information derived from ice core studies. An instrument to conduct large-area surveys of biomass density as a function of depth and microstructural features 
would enable significant progress in understanding algal  
dynamics in the rapidly changing polar marine environment.   The optical properties of sea water have been subject to multiple studies~(ex.~\cite{Gordon:76}). 
Efforts to determine the transport of light through ocean water have been undertaken.
Measurements are limited to specific locations and environments mostly close to the surface. There are few studies into optical properties at greater depths. 

This proceeding presents a conceptual design for an instrument for characterizing  the optical transport properties of glacial ice, sea ice, and the deep ocean, targeting the primary applications of neutrino telescope calibration and monitoring sea ice algal biomass. This multipurpose instrument would serve as a common calibration platform for all major neutrino telescopes as well as the climate science community.

\vspace{-.4cm}
\section{Instrument Design Drivers\label{sec_instrument}}
\vspace{-2.0ex}
The proposed instrument, the Deep Ice and Sea Calibration Observer (DISCO), combines two optical modalities: 1) ultra-sensitive cameras with precisely calibrated light sources that exploit the well-defined emission profiles and spectra to study optical properties through back-scatter imaging; 2) pulsed laser sources and precisely timed optical detectors that study absorption and scattering properties through correlated measurements of pulse intensity and timing. 

\begin{figure}
    \includegraphics[width=0.43\textwidth]{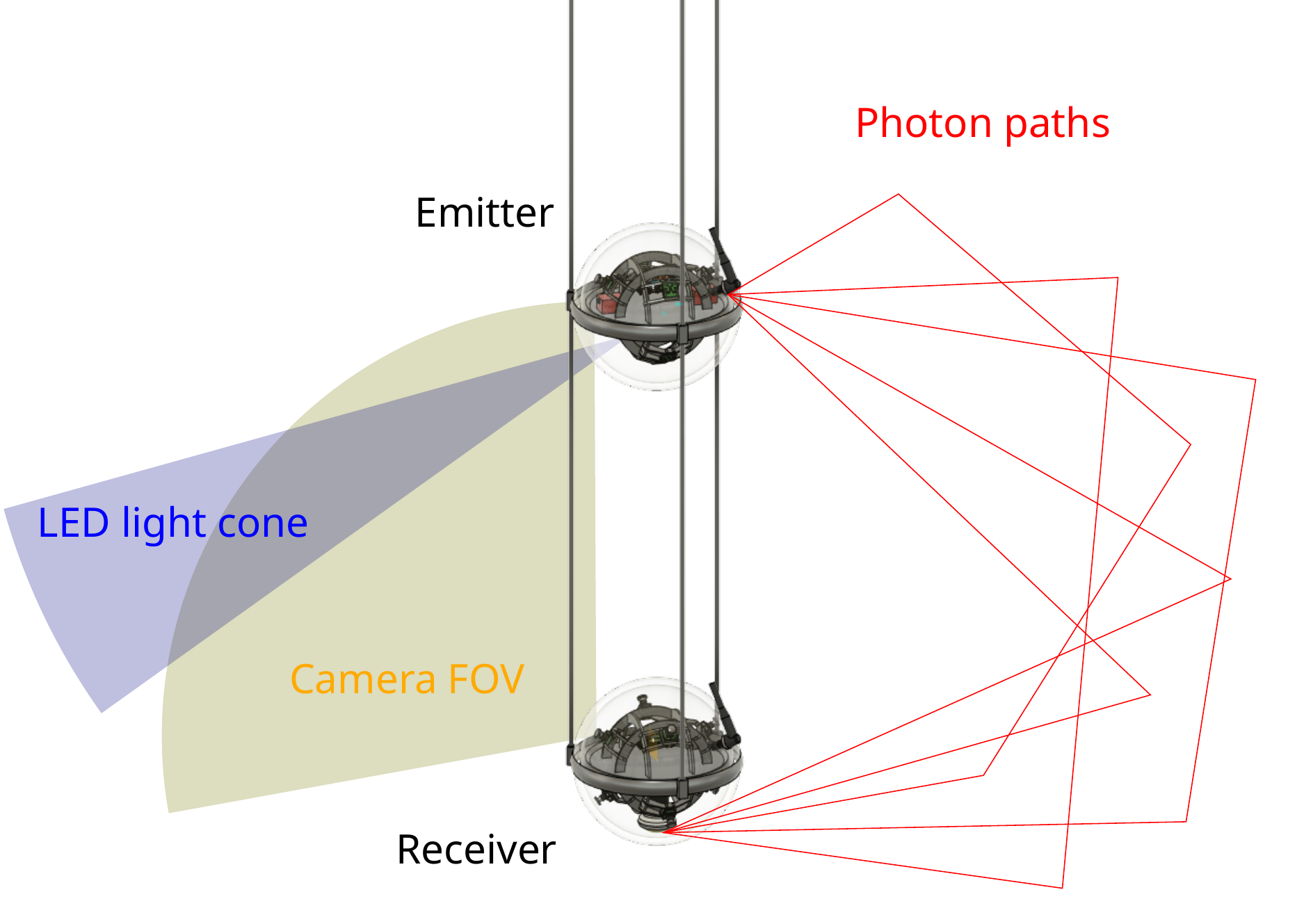} 
    \includegraphics[width=0.55\textwidth]{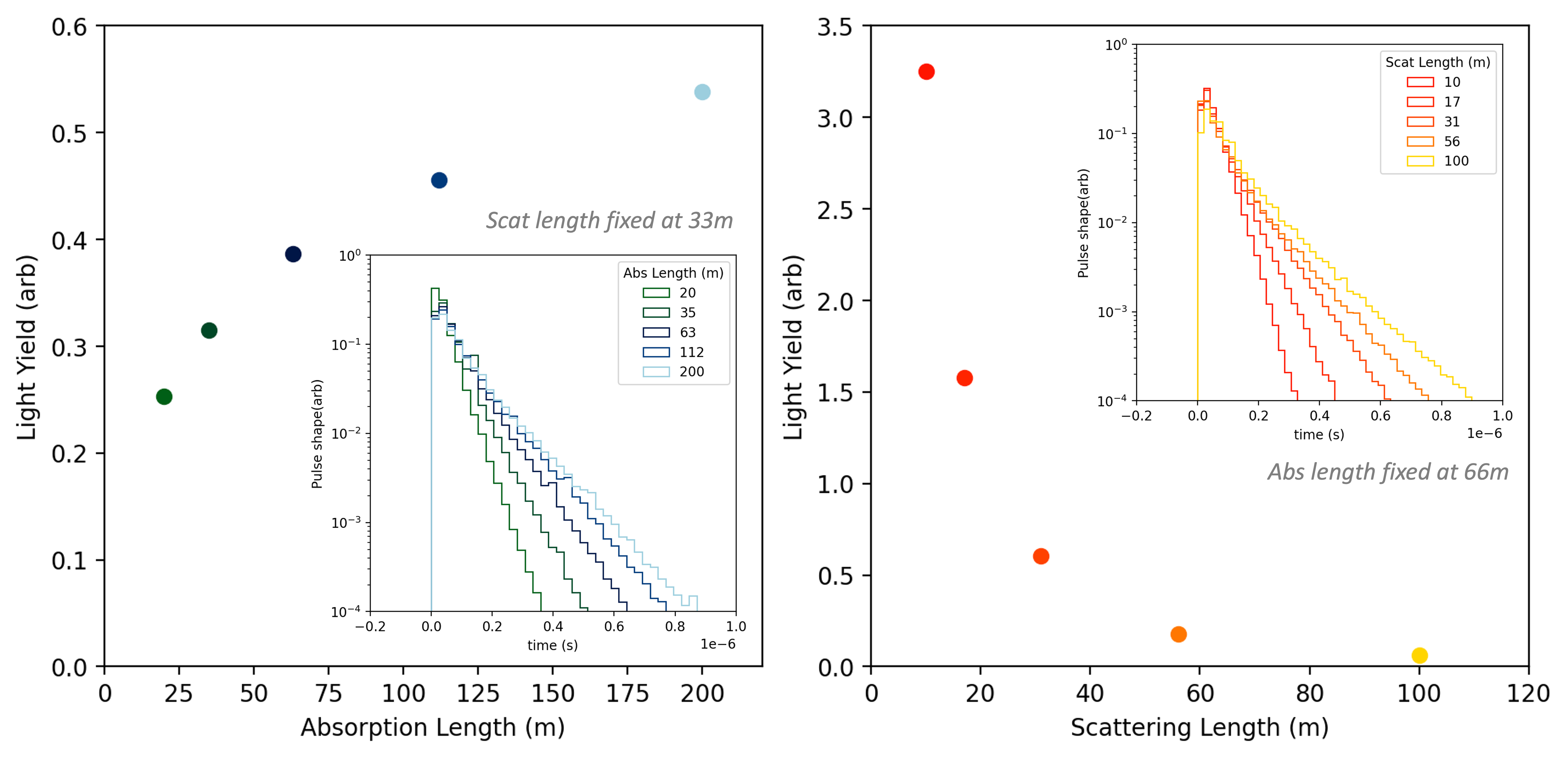} 
    \caption{Left: Concept of DISCO. LED light cone observed by cameras (left side) and Laser observed by the PMT logging system (right side). 
 Right: Absorption and scattering effects on pulse shape and returned light intensity for the pulsed laser measurement.}
     \label{fig:Outline}
    \label{fig:AbsScaFig}
 \end{figure}

The system (see Fig.~\ref{fig:Outline}) will consist of a receiver and emitter module, each contained in an independent pressure vessel and deployed via a winch system. With a robust design and enclosed within a borosilicate pressure housing, DISCO will be capable to withstand extreme environments in Antarctica boreholes or the deep sea. On-board monitoring systems maintain detailed logs of stability and allow for real-time gain adjustment that maximizes dynamic range. 

Three motivating applications are considered as Design Drivers: (1) Measurement of absorption and scattering coefficients as a function of location and depth in deep ocean water and glacial ice.  
(2) Determination of stratigraphy maps for ice tilt isochrones in ice-based observatories. 
(3) Quantification of biomass density and its correlation with local ice conditions within sea ice.  Complimentary applications include surveying equipment in the deep sea, studying birefringence, and measurement of bioluminescence and bioflourescence in ocean conditions. Still further  applications may be enabled, building upon the modular and expansible platform design.

{\bf Measurement of absorption and scattering coefficients:} Exploiting the physics capabilities of neutrino detectors requires precise modelling of optical properties of the detection medium. The medium is naturally occurring, so its properties are as-found and may be non-uniform or time-dependent, in the case of ice and water, respectively. Neutrino telescope observations rely on precise characterization of the in-medium properties; conversely, imperfectly characterized ice and water properties remain a major component of their systematic uncertainty budgets.

{\bf Stratigraphy maps to produce ice tilt isochrones for ice-based observatories: } The ice at the geographic South Pole is formed from compressed snow accumulated over more than 100,000 years~\cite{https://doi.org/10.1029/2000GL011351}. Atmospheric dust in the glacier reflects the air composition at the time of deposition, which is correlated to Earth temperature over geological time. The ice optical model at IceCube is comprised of optical absorption, scattering, and anisotropy parameters in each layer, assumed each to have been deposited at a fixed time (an isochrone).
If the glacier were stationary and the bedrock flat, the isochrones would be planar. However, glacial flow over the uneven continental surface has caused the isochronal layers to buckle and sheer. The 3D description of the isochronal surfaces is called the ``tilt map" and is a crucial ingredient input to the IceCube ice model. A further consequence of this glacial sheer is a birefringent crystal alignment that cause anisotropic propagation of light within the detector~\cite{abbasi2022situ}. 

The energy and angular reconstruction of IceCube events used in neutrino physics and astronomy analyses depends sensitively on the depth-dependencies within the ice model and also its tilt map.  Past attempts to measure the tilt used a device called the dust logger, which made laser based measurements in seven IceCube boreholes~\cite{bramall2005deep}. While highly valuable to IceCube,  (1) It did not have monitored or well-controlled gain of the photodetector, or intensity of the laser, which were adjusted by hand to avoid saturation, (2) there was insufficient information to independently quantify absorption and scattering, preventing validation of the universal description of ice within one layer.  Design driver 2 will provide precise stratigraphic / tilt data for future ice-based observatories,  
encompassing absolute absorption and scattering information within each isochrone and location-dependent anisotropy measurements.

{\bf 
Optical characterization of sea ice microstructure and 
algal biomass distributions in response to a warming climate:} Earth's sea ice covers form
vast frozen components of the climate system.
They serve as critical habitats for marine life,
from microbial communities living inside the ice,
to charismatic megafauna like penguins and polar bears
whose diets can be largely traced back to the algal
stocks hosted by sea ice.
Climate change, however, has significantly affected
the polar regions,
with the precipitous loss of Arctic sea ice
being one of the most visible
large-scale changes on Earth's surface connected to 
planetary warming.  Expanding our capabilities of  
studying the rapidly changing 
polar marine environment 
is critical to improving predictive models 
of climate change and the response of the 
cryosphere and the ecosystems it supports.

In sea ice there are brine and air 
inclusions, microbial organisms, particulates like dust or sediments, 
and larger-scale features like air pockets,
cracks, gap layers, and tree-like channels that 
drain melt ponds and enable fluxes 
of brine, sea water, and nutrients. They
all affect how light interacts with the ice (ex.~\cite{Warren_PTRSA_2019}).
 In particular,
there is significant interest in  
mapping the \textit{in situ} distribution of ice algal biomass, 
for which efficient, non-invasive methods remain sparse \cite{Cimoli:SciRep:2020}.  
DISCO will 
assess spatial variability 
in algal biomass and determine what
factors control their distributions, such as
brine porosity and connectedness, crystalline structure, 
and melting history.
For example, measuring
the vertical movement of algae  
to balance opposing light and nutrient resource gradients, 
correlated with local conditions,
may shed light on optimal strategies for survival 
in extreme, ephemeral, rapidly changing environments. 

To measure biomass concentration we will exploit its 
characteristic absorption spectrum \cite{Arrigo:E:2014}.
Absorptivity peaks strongly at 670~nm and has another broad feature in the range 400-500~nm with little absorption in between. This characteristic signature of chlorophyll \textit{a} 
can be used to separate biological spectral data 
from those of inorganic particulates
\cite{Light_JGR_1998,
Marks:TC:2017},
given multi-color illumination and imaging.
Radiative transfer models \cite{Arrigo_JGR_1991,Light_JGR_2004}
and related methods \cite{Cimoli:SciRep:2020}
can be used to invert 
measured optical signatures for microstructural and
biomass data.
Moreover, the analytic continuation method
for obtaining bounds on the effective complex
permittivity of composites, and inverse bounds on
microstructural parameters from bulk electromagnetic measurements,
has seen tremendous success in many applications, 
including
sea ice \cite{Golden_NAMS_2020}.

\vspace{-.4cm}
\section{Instrument Description and Measurement Principles}
\vspace{-.2cm}

\begin{figure}[tb]
    \centering
    \includegraphics[width=0.99\textwidth]{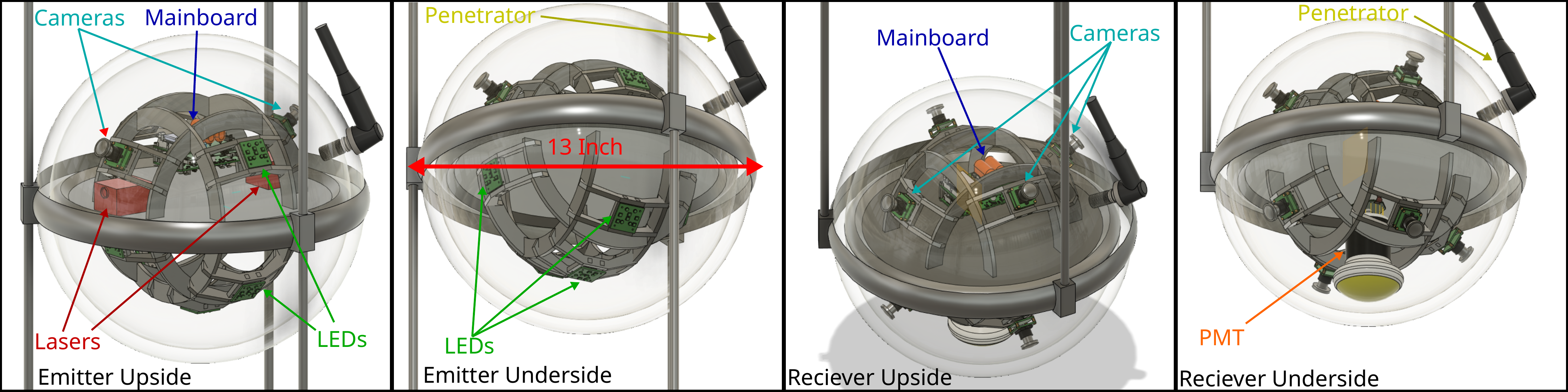} 
    \caption{A 3D model of the basic design concept for the emitter and receiver module.} 
    \label{fig:Tech_schem_1}
\end{figure}

DISCO consists of two spherical modules (Fig.~\ref{fig:Tech_schem_1}) that will contain several fixed cameras and illumination sources for the imaging system on a well-balanced inner mechanical structure for stable operations.  For the pulsed system, the top module will serve as emitter with two pulsed lasers, and the bottom as receiver, with a large photocathode area photomultiplier tube (PMT) pointing downwards, and an optically isolating baffle around its waist to shield the PMT from direct- or single-scattered laser light. Data storage and camera readout electronics will be present in each module, as well as control and acquisition systems for the laser and PMT. A communication channel between the two devices ensures synchronization \& commanding.

An electromechanical cable connects the system securely to the surface and is unwound with a winch system with passive heave compensation to stabilize and adjust the deployment depth. Modules are connected via a harness with adjustable separation. For near surface sea ice measurements or surveying activities, an independent operation offers flexibility in deployment configurations. 

\begin{figure}[tb]
    \centering
    \includegraphics[width=0.57\textwidth]{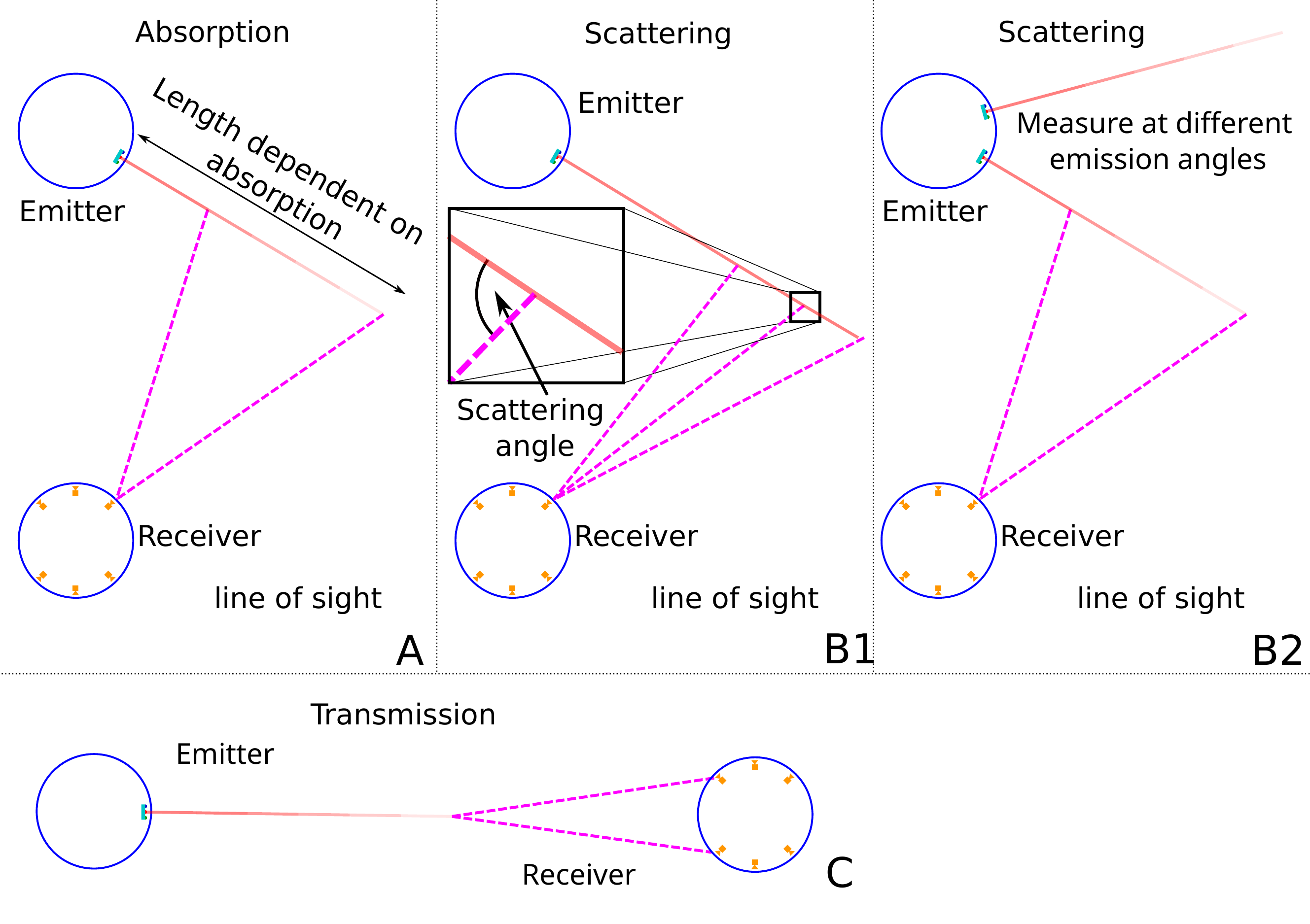} 
    \includegraphics[width=0.33\textwidth]{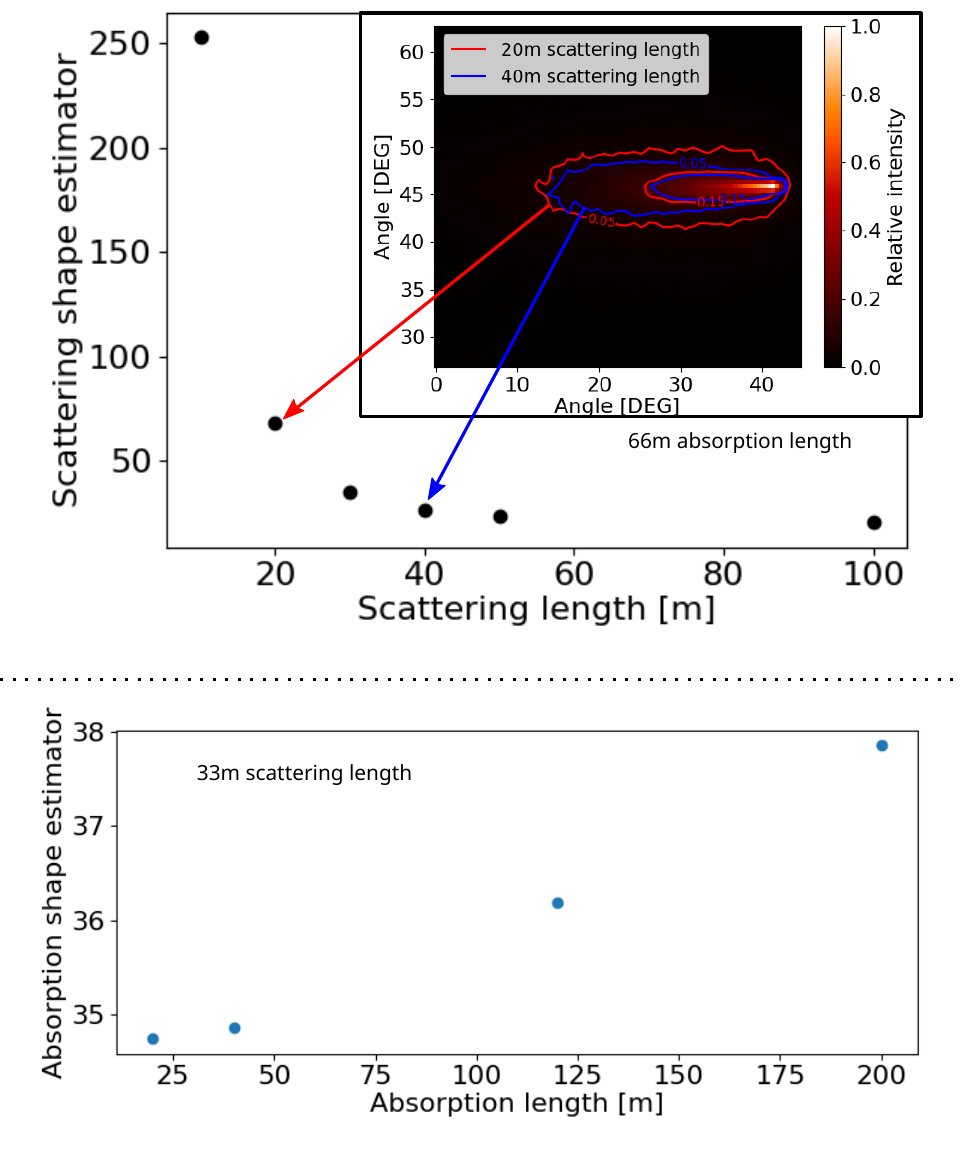} 
    \caption{Left: DISCO configurations for A: Absorption; B1: Scattering function B2: Anisotropy; C: Sea ice transmission measurement, Right: Expected camera image using a $2^{\circ}$ beam at 3~m distance and obtained optical medium parameters based on image.} 
    \label{fig:Tech_schem_2}
    \label{fig:LED}
\end{figure}

The camera system will access information on absorption and scattering strengths and the angular dependence of the scattering function by imaging the size and shape of scattered beams. Simulations to determine expected light signatures of illumination LEDs observed by the reciever module cameras have been conducted 
(Fig.~\ref{fig:LED}).
Since scattering and absorption have different wavelength dependencies, multichromatic measurements will aid disentangling optical medium properties. Optical transport anisotropy can be accessed by comparing the directional response.

The scattering measurement builds on well established experimental principles \cite{AVRORIN2012186},
but offers several additional benefits. It can measure the scattering function \textit{in situ} at greater depths and determine the wavelength dependence of the scattering function. Furthermore, the instrument is not designed for one specific environment and can hence be used for cross-calibration efforts of different media. Images carry significant information, leading to over-constrained problems which can be used to evaluate measurement uncertainties. As an improvement over previous surveying activity at Lake Baikal 
\cite{AVRORIN2012186}, DISCO will 
be more sensitive to backscattered light and potentially 
distinguish between scattering processes using polarized light. 

The laser system uses lower photon counts and fast pulsing to disentangle the effects of absorption and scattering behaviours through timing. Preliminary simulations suggest that absorption and scattering can be independently measured by comparing pulse shapes and intensities. Figure~\ref{fig:AbsScaFig} shows preliminary simulations of pulse shapes and intensities in homogeneous and isotropic media with different absorption and scattering lengths, in the ranges of relevance to neutrino telescopes.

The detection time of a photon after a pulsed emission encodes its travel distance. Absorption acts to extinguish light rays exponentially with distance, so shorter attenuation lengths suppress the late-time tails of received light pulses, as well as reducing total intensity. Increasing the scattering length also reduces intensity since less photons return from the medium before being absorbed. On the other hand, increasing scattering length also extends the pulse tail, photons travel further on average before detection. The leading edge of the pulse is also sensitive to scattering length, since its value influences the distribution of the shortest few photon paths.

The sea ice application is shallow, so ambient light intensity is expected to be too bright for use of the PMT system.  However, the camera system can be used for both scattering based biomass assay and also direct imaging of the local environment around the module under white light illumination.   Biomass density will be measured by exploiting the characteristic spectral shape associated with the chlorophyll. The emitter will be outfitted with light sources at 450, 550, and 670~nm, and camera images processed with optical reconstruction tools will be used to determine absorption strength at each wavelength. Enhanced absorption at 450 and 670~nm relative to intermediate wavelengths allows the spectral separation of biological matter from sediments and other absorbing backgrounds.   Geometrical configurations with the emitter free-floating in a borehole and the receiver underwater beneath the ice sheet, and with the two separated horizontally rather than vertically for a transmission- rather than reflection-based measurement will be explored. 

\vspace{-.4cm}
\section{Description of the DISCO Instrumentation}
\vspace{-.2cm}

The optical sensory systems combine cameras and PMTs to detect light from Lasers and LEDs. Compact, high performance, low-noise, high-sensitivity cameras capable of video and long exposure photo capture will be used. Camera lenses with large aperture to optimize light collection and a resolution and field of view (FOV) informed by simulation will be used. 
Receiver module cameras will have overlapping FOV and provide combined 4$\pi$ coverage. The emitter module will be equipped with two cameras pointing alongside the lasers in the ecliptic.
The PMT system 
records waveforms, needed for a precise timing of the rising edge of the incoming photon signal to determine the time between the light pulse emission from the laser system and the arrival of early photons in the PMT.  Time resolution of tens of nanoseconds is required to optimize sensitivity to local optical parameters on the leading edge, with tails out to approximately two microseconds. 

To avoid interference between the PMT and light sources for the cameras the PMT will be outfitted with an optical high pass filter that cuts off any light with a wavelength below 520~nm. The lasers used for the PMT will have wavelengths above this threshold and the main light source used for the camera will use a wavelength of 470~nm or shorter.
To ensure well defined and stable spatial and spectral properties under the mechanical and thermal stress during operations the beam will be continuously monitored using a beam splitter and a calibrated photon detector.

Backscattering measurements calibrated to this local reference will allow extraction of absolute optical parameters from the detector medium, relaxing constraints on laser stability and temperature independence. We anticipate the auxiliary optical path needing to be hermetically sealed in a dry nitrogen environment to avoid condensation or icing at low temperatures.
A circuit to adjust brightness automatically will be fed with a signal split from the backscatter monitoring PMT fed into a local microcontroller.

\vspace{-.4cm}
\section{Conclusion}\label{sec:conclusions}
\vspace{-.2cm}

We have outlined a multipurpose instrument for the calibration of neutrino telescopes. DISCO will be designed to be compatible and easily adaptable to existing site infrastructure to ensure that field deployments only require minimal adjustments. 

\bibliographystyle{ICRC}
\bibliography{references}

\end{document}